\documentclass[a4paper,12pt,aps,superscriptaddress]{revtex4}

\usepackage{fullpage,epsf}
\usepackage{graphicx}
\usepackage{amsmath}

\newcommand{\phic}{\phi_{\rm c}}

\renewcommand{\phi}{\ensuremath{\varphi}}  

\begin{document}

\title{Condensation Transitions in a One-Dimensional Zero-Range Process with a Single Defect Site}
\author{A.\ G.\ Angel}
\author{M.\ R.\ Evans}
\affiliation{School of Physics, University of Edinburgh,
Mayfield Road, Edinburgh EH9 3JZ, UK}
\author{D.\ Mukamel}
\affiliation{Department of Physics of Complex Systems,
Weizmann Institute of Science, Rehovot, Israel 76100}
\date{\today}

\begin{abstract}
Condensation occurs in nonequilibrium steady states when a finite
fraction of particles in the system occupies a single lattice site.
We study condensation transitions in a one-dimensional zero-range
process with a single defect site.  The system is analysed in the
grand canonical and canonical ensembles and the two are
contrasted. Two distinct condensation mechanisms are found in the
grand canonical ensemble.  Discrepancies between the infinite and
large but finite systems' particle current versus particle density
diagrams are investigated and an explanation for how the finite
current goes above a maximum value predicted for infinite systems is
found in the canonical ensemble.
\end{abstract}

\maketitle

\section{Introduction}

Condensation transitions \cite{BrazJPhysMRE} have been observed in a
wide variety of nonequilibrium steady states \cite{OEC,MKB}.  Such a
transition occurs when a single lattice site contains a finite
fraction of particles of the system.  Condensation transitions are also
manifested in a different guise in driven diffusive systems, where
typically the particles obey hard-core exclusion.  There the
condensation appears as a transition from a freely flowing fluid phase
to a jammed phase \cite{OEC}. Such jamming transitions have been
reported, for example, in the context of traffic flow modelling
\cite{CSS}.

A simple lattice model known as the Zero-Range Process (ZRP)
\cite{Spitzer} illustrates an equivalence between condensation and
jamming transitions and provides a framework within which condensation
transitions may be analysed
\cite{BrazJPhysMRE,OEC,GSS}.
In the model particles hop between lattice sites with rates which
depend on the number of particles at the departure site. If the rates
decay to zero or if they decay slowly enough to a finite value as the
particle number is increased, one finds a condensation transition
whereby a single lattice site becomes macroscopically occupied.

The connection between condensation and jamming transitions was first
made in the context of the Bus Route Model \cite{OEC} where an
approximate description assigns hopping rates to buses that decay as a
function of the distance to the next bus ahead. If the decay is
suitably slow a jamming transition arises whereby the buses cluster
together into a jam and there is a large region of the bus route
devoid of buses.  This approximate description of the bus route model
maps exactly onto a ZRP where the buses correspond to the sites of the
ZRP and the number of bus stops in front of a bus corresponds to the
number of particles at a site of the ZRP. Phase separation in driven
diffusive systems may also be related to condensation transitions and
the ZRP has been used to formulate a general criterion for phase
separation in one dimension \cite{KLMST}.

In addition to the condensation mechanism in the ZRP described above
(condensation associated with hopping rates that are decreasing
functions of the occupation of a site) one can have condensation
associated with {\em disorder} \cite{ME96,KF96}. Here disorder
refers to different sites having different hopping rates. The latter
mechanism has been shown to be equivalent to Bose-Einstein
condensation \cite{ME96,ME97}.  A particularly simple system that
exhibits Bose-Einstein-like condensation is where there is one slow
defect site at which a condensate forms at high enough density.

In the present work we study a simple model, with a single defect
site, that exhibits features of both condensation mechanisms
(bus-route-like and disorder-induced) mentioned above. The steady
state can be solved and the critical density for condensation in the
thermodynamic limit can be predicted. However, finite systems show
significant deviations from these predictions. In particular, the low
density, fluid phase appears to continue to higher densities than
would be allowed in the thermodynamic limit.  This results in an
overshoot in the current, above the thermodynamic saturation value.
Related finite size effects have been reported in traffic flow models
\cite{BSSS}.

Our aim in this paper is to analyse in detail how the finite size
overshoot in the current occurs.  Now, since the ZRP dynamics
conserves particle number the natural ensemble to consider is the
canonical ensemble. However, one finds that analysis of the model is
more direct within the grand canonical ensemble (i.~e.\ where the
particle number fluctuates). It has been shown that in the infinite
system limit the two ensembles are equivalent \cite{GSS}.  However,
since we are interested in the behaviour of large but finite systems,
we broach the issue of the inequivalence of the two ensembles on
finite systems.  We are able to analyse the single defect site model
in both the canonical and grand canonical ensembles and find that they
give different predictions for the finite size overshoot in the
current.  In particular, within the canonical ensemble the overshoot
is predicted to peak at a density $\phi = \phi_{\mathrm{c}}
+O(M^{-1/2})$, where $\phi_{\mathrm{c}}$ is the thermodynamic critical
density and $M$ is the number of sites, and we confirm this by
simulations.

The paper is organised as follows: in section 2 we define the ZRP and
discuss the two types of condensation mechanism; in section 3 we
introduce the single defect site (SDS) model and analyse the
condensation transition within the grand canonical ensemble; in
section 4 we present exact numerical results and simulations that
illustrate the overshoot of the low density fluid phase on a finite
system in the canonical ensemble; in section 5 we analyse the SDS
model in the canonical ensemble; and we conclude in section 6.

\section{The zero-range process}
\subsection{Model definition}
\label{modeldefn}
We consider a one-dimensional, asymmetric zero-range process.
We have a one-dimensional lattice of $M$ sites labelled $\mu = 1
\ldots M$, upon which reside $L$ indistinguishable particles, each
site being able to hold any number ($0 \ldots L$) of particles.
Particles move from site to site under a dynamics which conserves the
number of particles and is described as follows: a particle from site
$\mu$ hops to site $\mu +1$ with a rate, $\mathrm{u}_{\mu}(n_{\mu})$,
which can depend on both the departure site, $\mu$, and the number of
particles on this site, $n_{\mu}$.  The boundary conditions are
periodic; a particle hopping from site $M$ will hop to site $1$.  The
configuration of the system at any given time is specified by the set
of occupancies of each site, $\left\{ n_{\mu} \right\}$.

Thus we consider the zero-range process on a finite system of size
$M$; in particular we shall be interested in the limit of large $M$
where the particle density $\phi = L/M$ is held fixed.

\subsection{Steady state}
The one-dimensional zero-range process, with particles hopping to
adjacent sites in one direction only, has a steady state which is
straightforwardly solved \cite{BrazJPhysMRE,Spitzer}.
Requiring that there be no net flow of probability to or from a
configuration, one finds the probability of a configuration, $\{ n_{\mu}
\}$, in the steady state to be
\begin{equation}
\mathrm{P}( \{ n_{\mu} \} ) = \frac{1}{\mathrm{Z}(M,L)} \prod_{\mu=1}^{M} 
\mathrm{f}_{\mu}(n_{\mu})\;,
\label{steadystateP}
\end{equation}
where the $\mathrm{f}_{\mu}$ are given by
\begin{equation}
\mathrm{f}_{\mu}(n) = \left\{
\begin{array}{cl}
\prod\limits_{m=1}^{n} \frac{1}{\mathrm{u}_{\mu}(m)} & \mbox{if $n \geq 1$} \\
1 & \mbox{if $n = 0$\;,}
\end{array}
\right.
\label{marginalsf}
\end{equation}
and $\mathrm{Z}(M,L)$ has been introduced to ensure the probabilities
are correctly normalised.  This normalisation is analogous to
the canonical partition function from equilibrium statistical mechanics, being
the sum of the un-normalised probability over all allowed
configurations,
\begin{equation}
\mathrm{Z}(M,L) = 
\sum_{n_{1}, n_{2}, \ldots, n_{M}} 
\delta \left( \sum_{\mu=1}^{M} n_{\mu} - L \right) 
\prod_{\mu=1}^{M} \mathrm{f}_{\mu}(n_{\mu})\;.
\label{bigZcanongeneral}
\end{equation}
Here we have summed over all occupancies for each site, from zero to
infinity, and included the delta function to ensure that only those
configurations with the correct total number of particles contribute.
Thus $Z(M,L)$ is the normalisation in the canonical ensemble.

This steady state probability distribution can be used to calculate
other steady state properties of the system.  For our purposes the
most important of these is the average hopping rate from a site, $v$,
which we refer to as the current.  In the steady state this is
independent of site and is given by
\begin{equation}
v = \frac{1}{\mathrm{Z}(M,L)}
\sum_{n_1} \mathrm{u}(n_1) 
\sum_{n_{2},  \ldots, n_{M}}
\delta \left( \sum_{\mu=1}^{M} n_{\mu} - L \right)  
\prod_{\mu=1}^{M} \mathrm{f}_{\mu}(n_{\mu}) =
 \frac{\mathrm{Z}(M,L-1)}{\mathrm{Z}(M,L)}\;,
\label{vequals}
\end{equation}
where in the final step we have used the fact that
$\mathrm{u}_{\mu}(0) = 0$ (the departure rate from an empty site is
zero) and $\mathrm{u}_{\mu}(x)
\mathrm{f}_{\mu}(x) =
\mathrm{f}_{\mu}(x-1)$ for $x \geq 1$.

\subsection{Condensation transition - grand canonical analysis}
\label{CondGCEsec}
Depending on the chosen hopping rates the system can exhibit a
condensation transition in the thermodynamic limit $M \to \infty$ with
the density of particles fixed. In this case, at low density the
system is in a {\em fluid} phase, by which we mean that all sites have
average occupancies that are infinitesimal fractions of the total
number of particles.  On the other hand, at high density the system is
in a {\em condensed} phase wherein a single site holds a finite
fraction of the total number of particles.  We shall show that there
are two distinct mechanisms by which this may occur.

The simplest approach to analysing the condensation transition is to
use the grand canonical ensemble.  The functions $ f_\mu(n)$ given in
(\ref{marginalsf}) are only defined up to some multiplicative factor,
$z^n$, which we have tacitly set to one. Reinstating $z$, we can
interpret $z$ as the fugacity.  Then we can get the equivalent of the
grand canonical partition function in the usual way \cite{khuangbook},
by summing over all configurations, including those that violate the
constraint on the total particle number, and then choose $z$ such that
the average total particle number is equal to $L$:
\begin{eqnarray}
{\cal Z}(M,z) &=& \sum_{n_{1}=0}^{\infty} \sum_{n_{2}=0}^{\infty} \dotsm
\sum_{n_{M}=0}^{\infty} \; \prod_{\mu=1}^{M} z^{n_\mu}
\mathrm{f}_{\mu}(n_{\mu})\label{Zgce}\\
&=& \prod_{\mu=1}^{M}
\mathrm{F}_\mu(z)\\
\mbox{where}\quad \mathrm{F}_\mu(z) &=& 
\sum_{n=0}^{\infty} z^n
\mathrm{f}_\mu(n)\;.
\label{bigF}
\end{eqnarray}
The constraint on the average particle number is written as follows
\begin{equation}
\sum_{\mu = 1}^{M} \left< n_{\mu} \right> = L\;,
\label{summuLeq}
\end{equation}
with the average occupation of each site, $\left< n_{\mu} \right>$,
being found as usual by:
\begin{equation}
\left< n_{\mu} \right> = 
\frac{z\,\partial \ln \mathrm{F}_\mu(z)}{\partial z}\;,
\label{nmueq}
\end{equation}
which yields
\begin{equation}
\phi = \frac{z}{M} \sum_{\mu=1}^{M} \frac{\mathrm{F}'_\mu(z)}{\mathrm{F}_\mu(z)}
\label{saddlepoint1}
\end{equation}
where $\phi$ is the particle density $L/M$.

\subsubsection{Different types of condensation}
\label{mechsec}
To see the two apparently different mechanisms we analyse the
behaviour of the rhs of (\ref{saddlepoint1}) as $z \to z_{\rm max}$
where $z_{\rm max}$ is the radius of convergence of
(\ref{bigF}).
As $z \to z_{\rm max}$ one of two things can happen, either the rhs of
(\ref{saddlepoint1}) will converge, or it will diverge.

If the rhs of (\ref{saddlepoint1}) converges at $z = z_{\rm max}$,
then this will give a critical density, $\phi_{\mathrm{c}}$, above
which (\ref{saddlepoint1}) can no longer be satisfied. Hence there is
a phase transition, whereby the excess density condenses
onto a single site.  We refer to this as mechanism A\@.

If the rhs of (\ref{saddlepoint1}) diverges at $z = z_{\rm max}$, then
it is still possible to have a phase transition by a 
different mechanism.  Recall that each term in the sum in
(\ref{saddlepoint1}) is the density of particles
at a site.  It is possible for the rhs of (\ref{saddlepoint1}) to
diverge due to only a single term, say site 1, of the sum diverging.
This would imply that site 1 contains
a finite fraction of the total number of particles, i.~e.\ we have a
condensate.  We refer to this as mechanism B\@.  To see that mechanism B
is reminiscent of the Bose-Einstein condensation mechanism, one thinks
of the particles as bosons and the sites of the ZRP as Bose states.
Then site 1 into which condensation may occur corresponds to the
ground state of the bosonic system.

The choice of hopping rates will determine which mechanism takes
place.  Previously studied cases are: 
\begin{enumerate}
\item The hopping rates are the
same for each site and decay with increasing particle number
\cite{BrazJPhysMRE,OEC}; then we can have a condensation through
mechanism A\@.  
\item The hopping rates do not depend on the number of
particles at a site, but do depend on the site, i.~e.\ the system is
disordered \cite{BrazJPhysMRE}.  Then we can have a condensation
through mechanism B\@. One very simple realisation of this is a single
defect site with a hopping rate less than the other sites.  Then at
high density this site will support a condensate.
\end{enumerate}

One important difference between mechanisms A and B is that in A there
can be a spontaneous symmetry breaking, in that the site supporting
the condensate is selected at random. With mechanism B, on the other
hand, the term in the sum which diverges specifies which site supports
the condensate (taken as site 1 above).

If we have a mixture of decay and disorder in the hopping rates, then
it is possible to have a phase transition through either mechanism. In
the present work we investigate this possibility. 

\subsubsection{Overshoot in the current $v$}
In the grand canonical ensemble the current is given by
\begin{eqnarray}
v & = & \left< u_{1} \right> \nonumber \\
& = & \frac{1}{{\cal Z}(M,z)} \sum_{n_{1}=0}^{\infty} \mathrm{u}_{1}(n_{1}) \mathrm{f}_{1}(n_{1}) z^{n_{1}} \sum_{n_{2},n_{3}, \dots, n_{M} = 0}^{\infty} \prod_{\mu=2}^{M} \mathrm{f}_{\mu}(n_{\mu}) z^{n_{\mu}} \nonumber \\
& = & z\;,
\label{v=z1}
\end{eqnarray}  
where to proceed to the final line we have used the fact that
$\mathrm{u}_{\mu}(0) = 0$ and $\mathrm{u}_{\mu}(x)
\mathrm{f}_{\mu}(x) = \mathrm{f}_{\mu}(x-1)$ for $x \geq 1$.

Thus in the grand canonical ensemble we have an equivalence between
the current, $v$, and the fugacity, $z$.  From the discussion of the
behaviour of $z$ in the previous subsection we deduce that, in the
{\em infinite} system limit, $v$ increases with density in the fluid
phase until at density $\phi_{\rm c}$ the current saturates at $v =
z_{\rm max}$ and we enter the condensed phase.

However, on a large but {\em finite} system, things are more subtle.
In mechanism B one can always solve (\ref{saddlepoint1}) for $z<
z_{\rm max}$. Thus $v$ should always be less than its saturation value
$z_{\rm max}$. In mechanism A, on the other hand, one cannot satisfy
(\ref{saddlepoint1}) for $\phi >\phic$ with $z\leq z_{\rm max}$. Thus
it is possible that $v$ overshoots the value $z_{\rm max}$ \cite{KMP}. In
simulations (in the canonical ensemble) of finite systems an overshoot
in the current versus density diagram can be observed, i.~e.\ the
current appears to go above the expected maximum value before dropping
back down at higher density (see section~\ref{sec:numerics}).  Thus on
a finite system there can appear to be a continuation of the fluid
phase into the region where, on the infinite system, one would expect
a condensate.

\section{Single Defect Site}
We now restrict ourselves to a very simple ZRP with all hopping rates
constant and equal, except for a single slow defect site which has a
hopping rate that decreases for increasing particle number. We refer
to this as the single defect site (SDS) model. 
We begin by analysing the model within the grand canonical ensemble.

We consider the one-dimensional lattice of $\mu = 1 \ldots M$ sites
with periodic boundary conditions from section \ref{modeldefn} and now
choose the following hopping rates (for $l > 0$)
\begin{eqnarray}
\mathrm{u}_{1}(l) & = & p \left( 1 + \displaystyle \frac{\lambda}{l} \right)
\label{simplehop1}\\[1ex]
\mathrm{u}_{\mu }(l) & = & 1\quad\quad\mbox{for}\quad \mu>1\;, 
\label{simplehopmu}
\end{eqnarray}
with $p<1$ and we generally take $p \leq 1/(1+\lambda)$, giving site 1 a hopping rate slower than all the others for occupancies $l \geq 1$.
Then from (\ref{marginalsf})
\begin{eqnarray}
\mathrm{f}_1(n)&=& \frac{1}{p^n} \ \frac{n!}{(1+\lambda)_n} 
\label{simplef1}\\[1ex]
\mathrm{f}_{\mu}(n) &=& 1\quad\quad\mbox{for}\quad \mu>1\;, \label{simplefmu}
\end{eqnarray}
where $(a)_n$ is the Pochhammer symbol defined by
\begin{eqnarray}
(a)_{m} & \equiv & a(a+1)(a+2) \dots (a+m-1) \nonumber \\
(a)_{0} & \equiv & 1\;.
\end{eqnarray}
Also, from (\ref{bigF}) we have
\begin{eqnarray}
\mathrm{F}_1(z)&=& \sum_{n=0}^{\infty} \left(\frac{z}{p}\right)^n
 \frac{n!}{(1+\lambda)_n}  = {}_{2}\mathrm{F}_{1}(1,1\,;1+\lambda\,;
z/p) 
\label{bigF1stih} \\
\mathrm{F}_{\mu}(z) &=& 
\mathrm{F}(z)=\frac{1}{1-z}
\quad\mbox{for}\quad \mu>1\;,
\end{eqnarray}
where the hypergeometric function \cite{AAR} is defined as
\begin{equation}
{}_{2}\mathrm{F}_{1}(a,b\,;c\,;x) 
= \sum_{m=0}^{\infty} \frac{ (a)_m\, (b)_m }{(c)_m}\frac{x^m}{m!}\;.
\end{equation}
To see the equivalence in (\ref{bigF1stih}) note that $(1)_{m} = m!$.

If we look at the grand canonical analysis, specifically equation
(\ref{saddlepoint1}), for the SDS model the slow site gives a ratio of
hypergeometric functions, while the `non-slow' sites, $\mu = 2 \ldots
M$, give a ratio of two simple geometric sums
\begin{eqnarray}
\phi & = & \frac{z}{M} \left(  
\frac{\mathrm{F}_{1}^{\prime}(z)}{\mathrm{F}_{1}(z)} + 
(M-1) \frac{\mathrm{F}^{\prime}(z)}{\mathrm{F}(z)}
   \right) \nonumber \\[1ex]
& = & \frac{z}{M}\frac{ {}_{2}\mathrm{F}_{1}(2,2\,;2+\lambda\,;z/p) }
{ p(1+\lambda) \; {}_{2}\mathrm{F}_{1}(1,1\,;1+\lambda\,;z/p) } 
+ \left( 1 - \frac{1}{M} \right) \frac{ z }{ 1-z }  \\[0.5ex]
&=& \frac{\langle n_1(z)\rangle}{M} +
\left( 1 - \frac{1}{M} \right) 
\langle n(z)\rangle\;.
\label{saddlepointsimple}
\end{eqnarray}
In the last line $\langle n_1(z)\rangle$ means the 
average number of particles  at site 1,
whereas $\langle n(z)\rangle$ means the average
number of particles at a site  $\mu \neq 1$.
Thus, as $M \rightarrow \infty$, the critical density will be
\begin{equation}
\phi_{\mathrm{c}} = \frac{p}{1-p}\;.
\label{phiC}
\end{equation}
To see this we take the limit $z \to z_{\mathrm{max}}=p$
and $M\to \infty$ of the
expression (\ref{saddlepointsimple}) such that
$\langle n_1 \rangle/M \to 0$. This last condition
ensures we are not in the condensed phase.
However, for $ \phi > \phi_{\rm c}$
we must have $\langle n_1(z)\rangle/M$ finite to solve
(\ref{saddlepointsimple}),
thus we have a condensate on the slow site.


In order to determine  the finite size corrections
to the current $v=z$  we  analyse (\ref{saddlepointsimple})
for $M$ large but finite.
We  invoke the following results \cite{AAR}
\begin{eqnarray}
\mbox{if}\quad c > a+b
\quad \lim_{x\to 1^-} {}_{2}\mathrm{F}_{1}(a,b\,;c\,;x)
&=& \frac{ \Gamma(c)\,\Gamma(c-a-b)}{\Gamma(c-a)\,\Gamma(c-b)}\\[1ex]
\mbox{if}\quad c = a+b
\quad \lim_{x\to 1^-} \frac{{}_{2}\mathrm{F}_{1}(a,b\,;c\,;x)}{
|\ln(1-x)|}
&=& \frac{ \Gamma(a+b)}{\Gamma(a)\,\Gamma(b)}\\[1ex]
\mbox{if}\quad c < a+b
\quad \lim_{x\to 1^-} \frac{{}_{2}\mathrm{F}_{1}(a,b\,;c\,;x)}{
(1-x)^{c-a-b}}
&=& \frac{ \Gamma(c)\,\Gamma(a+b-c)}{\Gamma(a)\,\Gamma(b)}\;
\end{eqnarray}
where $\Gamma(x)$ is the usual Gamma function.
Thus
\begin{eqnarray}
\mbox{if}\quad \lambda > 2&&
 \lim_{z\to p^-} 
\langle n_1(z)\rangle = \frac{1}{(\lambda-2)}\\[1ex]
\mbox{if}\quad \lambda = 2&&
 \lim_{z\to p^-} 
\frac{\langle n_1(z)\rangle}{|\ln(1-z/p)|} = 1\\[1ex]
\mbox{if}\quad 1<\lambda < 2&&
\lim_{z\to p^-} 
\langle n_1(z)\rangle(1-z/p)^{2-\lambda}
= \frac{(\lambda-1)^2\, \pi}{\sin(\pi(\lambda-1))}\label{lambda22}\\[1ex]
\mbox{if}\quad \lambda = 1&&
 \lim_{z\to p^-} 
\langle n_1(z)\rangle(1-z/p)|\ln(1-z/p)| = 1\\[1ex]
\mbox{if}\quad \lambda < 1&&
 \lim_{z\to p^-} 
\langle n_1(z)\rangle(1-z/p)
=  (1-\lambda)\;,
\end{eqnarray}
where to obtain (\ref{lambda22}) we have used the identity
\begin{equation}
\Gamma(x) \Gamma(1-x) = \frac{\pi}{\sin(\pi x)}\;.
\end{equation}
Therefore when $\phi >\phi_{\mathrm{c}} = p/(1-p)$
for $\lambda \leq 2$ one can satisfy
(\ref{saddlepointsimple}) by choosing
\begin{eqnarray}
\mbox{if}\quad \lambda = 2&& \label{zgc1}
1-z/p\sim \exp\left[-M(\phi-\phic)\right]\\[1ex]
\mbox{if}\quad 1<\lambda < 2&&
1-z/p \sim 
\left[ \frac{ (\lambda-1)^2 \pi}{\sin(\pi(\lambda-1)) (\phi-\phic)}\frac{1}{M}
\right]^{1/(2-\lambda)}\\[1ex]
\mbox{if}\quad \lambda = 1
&& 1-z/p \sim \frac{1}{\phi-\phic}\frac{1}{M\ln M}\\[1ex]
\mbox{if}\quad \lambda < 1&&
1-z/p \sim \frac{(1-\lambda)}{\phi-\phic}\frac{1}{M}\;.
\label{zgc4}
\end{eqnarray}
For $\lambda >2$ on the other hand,
we cannot satisfy (\ref{saddlepointsimple}) 
as $z\to p^-$ when $\phi >\phi_{\mathrm{c}}$.
Instead we must consider $z >p$.
Thus the grand canonical treatment implies
an overshoot in the current $v$ ($=z$) for $\lambda > 2$.

In order to consider $z>p$ we impose cut-offs $n_\mu =L$ in the sums
of (\ref{Zgce}).  This explicitly ensures that no site contains more
particles than the total number $L$.  Then
\begin{eqnarray}
\langle n_1(z) \rangle
= \frac{z}{p(1+\lambda)}\, \frac{
{}_{2}\mathrm{F}^{(L)}_{1}(2,2\,;2+\lambda\,;
z/p) }{
{}_{2}\mathrm{F}^{(L)}_{1}(1,1\,;1+\lambda\,;
z/p) }
\label{n1co}\\
 \mbox{where}\quad
{}_{2}\mathrm{F}^{(L)}_{1}(a,b\,;c\,;x) 
= \sum_{m=0}^{L} \frac{ (a)_m\, (b)_m }{(c)_m}\frac{x^m}{m!}\;.
\label{limitL}
\end{eqnarray}
For $\lambda > 2$ we write $z/p$ in  the following form:
\begin{equation}
\frac{z}{p} = \exp(\alpha (L)).
\end{equation}
To satisfy (\ref{saddlepointsimple}) for $\phi >\phic$ we require
$\langle n_1(z) \rangle = {\cal O}(L)$.  It turns out that as $z$
increases above $p$ this is first achieved when the numerator of
(\ref{n1co}) is ${\cal O}(L)$ and the denominator ${\cal O}(1)$. In
anticipation of this we write
\begin{equation}
\langle n_1 \rangle \simeq
 \frac{\mathrm{G}_{\lambda-1}}{A+ \mathrm{G}_{\lambda}}
\quad \mbox{where}\quad
\mathrm{G}_\gamma = 
\int_{1}^{L} \frac{\exp(k \alpha(L))}{k^{\gamma}} \,\mathrm{d} k 
\end{equation}
and $A = {\cal O}(1)$, i.~e.\ we have replaced the sums in
(\ref{n1co},\ref{limitL}) by integrals over the asymptotic forms of the
summands, retaining a finite error term $A$ in the denominator.
Then we require that $\mathrm{G}_\lambda = {\cal O}(1)$ and
$\mathrm{G}_{\lambda-1} = {\cal O}(L)$.  Now the asymptotic behaviour
of $\mathrm{G}_\gamma$ is
\begin{equation}
\mathrm{G}_\gamma \sim \frac{\exp(L\alpha(L))}{ L^\gamma \alpha(L)}
\quad \mbox{for}\quad L\alpha(L) \gg 1\;.
\end{equation}
Thus we require
\begin{equation}
\frac{\exp(L\alpha(L))}{ L^\lambda \alpha(L)} = {\cal O}(1)\;,
\end{equation}
which implies
\begin{equation}
\alpha = (\lambda -1) \frac{\ln L}{L} +
\frac{\ln(\ln L)}{L} + {\cal O}(1/L)\;.
\end{equation}

We can now summarise the results of the grand canonical analysis
for a large but finite  system
\begin{itemize}
\item[{\bf A}]
For $\lambda > 2$, $\langle n_1 \rangle$ remains
finite as $z \to z_{\mathrm{max}}$ and we identify this case with
condensation mechanism A of section \ref{mechsec}.
In the condensed phase,
in order to satisfy (\ref{saddlepointsimple})
$z=v$ overshoots its saturation value $p$.
The approach of $v$ to
its saturation value as particle number $L$
increases is $v/p \simeq 1+ (\lambda-1)\ln L/L$.

\item[{\bf B}]
For $\lambda \leq 2$, we can access the condensed phase through
(\ref{saddlepointsimple}) and $v$ approaches its saturation value from
below as detailed in (\ref{zgc1}--\ref{zgc4}).  We identify the case
$\lambda \leq 2$ with mechanism B\@.
\end{itemize}

\section{Simulations and exact numerics in the canonical ensemble}
\label{sec:numerics}
In this section we will present results obtained from Monte Carlo
simulations of the system and exact numerical calculations of
$\mathrm{Z}(M,L)$ in the canonical ensemble (fixed particle number).
What we shall find is that on finite systems the overshoot in the
current, $v$, is clearly present whenever $\lambda >0$. This is in
contradiction to the grand canonical analysis of the previous section
which predicts an overshoot only for $\lambda>2$.

To observe the overshoot in the current, $v$, exact numerical
calculations of $\mathrm{Z}(M,L)$ for systems of sizes up to $M=1000$,
with up to $L=2000$ particles were undertaken.  These calculations are
performed recursively
\cite{OEC}.  It is then possible to
obtain $v$ via the relation (\ref{vequals}).

\begin{figure}[htb]
\begin{center}
\includegraphics[angle=270,width=10cm]{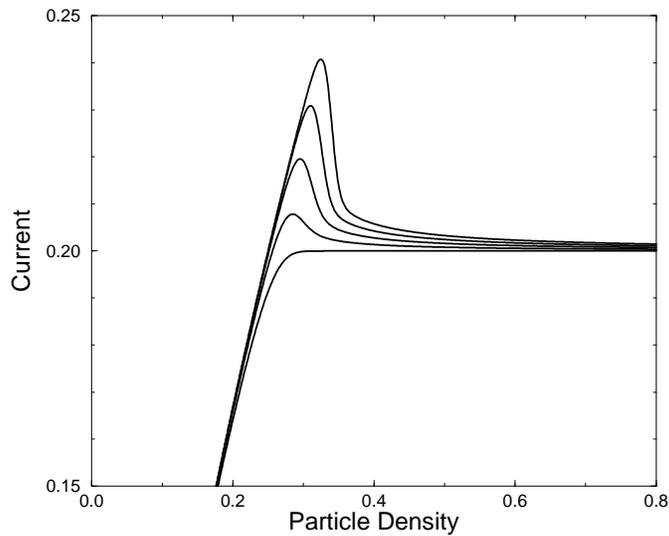}
\caption{
Exact numerical calculation of the current, $v$, plotted against
particle density for a system of size $M=1000$, with $p=0.2$ and
$\lambda = 0,1,2,3,4$ from bottom to top.
\label{vdifflambdafig}
}
\end{center}
\end{figure}

The current, $v$, is plotted against particle density at fixed system
size for various values of $\lambda$ and $p=0.2$ in
Figure~\ref{vdifflambdafig}.  In this figure we see that for low
density the current increases whereas for high density it saturates.
For $\lambda >0$, in between these two regimes is an overshoot in
the current where we observe non-monotonic behaviour.  Remember that,
for an infinite system, as the density is increased we would expect
to see $v$ increase to a maximum value and stay there, yielding a
current versus density curve that is non-analytic at the critical
density.  Thus, for the infinite system, when $v$ is increasing the
system is in a fluid phase, and when $v$ saturates the system is in a
condensed phase.  From Figure~\ref{vdifflambdafig} we see that on a
{\em finite} system, for $\lambda >0$, $v$ overshoots above its
expected saturation value.  This effect increases with increasing
$\lambda$. 

\begin{figure}[htb]
\begin{center}
\includegraphics[angle=270,width=10cm]{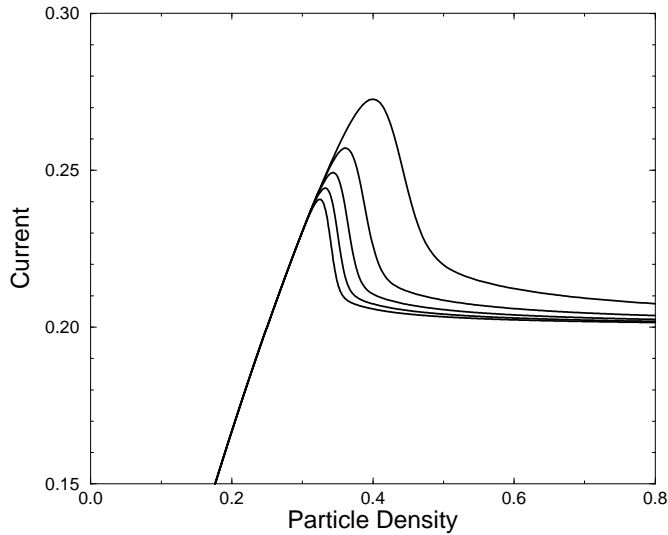}
\caption{
Exact numerical calculation of the current, $v$, plotted against
particle density for a system with $\lambda=4$, $p=0.2$ and varying
system size $M=200, 400, \ldots, 1000$ from top to bottom.
\label{vdiffMfig}}
\end{center}
\end{figure}

Notice also that for a finite system, $v$ crosses its predicted
saturation value for an infinite system, in this case $p$, close to
the predicted critical density for an infinite system, given by
$p/(1-p)$ from equation (\ref{phiC}).  The current, $v$, is also
shown in Figure~\ref{vdiffMfig}, plotted against particle density,
but for different system sizes, $M=200, 400, \ldots, 1000$ with $\lambda
= 4$ and $p=0.2$, from which it can clearly be seen that the overshoot
decreases in severity for increasing system size.  Thus the current
versus density curve will approach the expected infinite system
result.

\begin{figure}[htb]
\begin{center}
\includegraphics[angle=0,width=12cm]{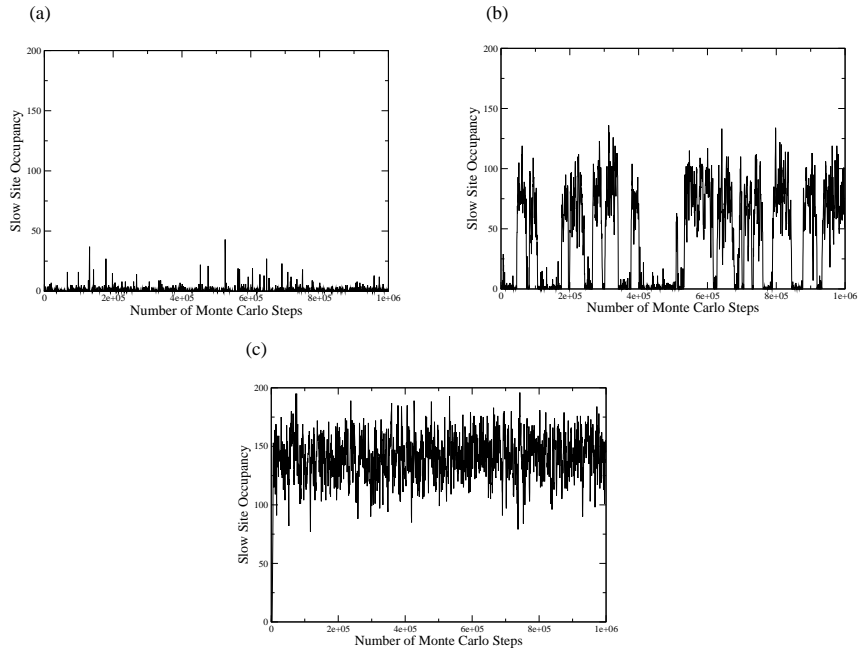}
\caption{
Graphs showing the number of particles on the slow site against time
for $\lambda = 4$, $p=0.2$ with $M=1000$, obtained via Monte Carlo
simulation.  (a) System is at $\phi = 0.29$, between expected critical
density, in this case $0.25$, and maximum of the overshoot.  This is
an extension of the fluid phase.  (b) System is at $\phi = 0.34$, just
past the maximum of the overshoot.  This is the coexistence region, a
small condensate is seen to appear intermittently, but is not stable
over long periods of time. (c) System is at $\phi = 0.40$, beyond the
`end' of the overshoot region where the current is approaching its
expected maximum value $v=p$.  Here the large pile of particles at the
slow site is stable and the system is in the condensed phase.
\label{3animouts}
}
\end{center}
\end{figure}

In order to investigate what is actually happening in the overshoot
region, Monte Carlo simulations of the system were run at chosen
points within the overshoot.  
The simulations were run in three key areas: firstly, between the
predicted critical density on an infinite system (\ref{phiC})
and the maximum of the overshoot; secondly, between the maximum of the
overshoot and where $v$ drops down close to its expected maximum value
of $p$; and thirdly where $v$ has dropped down close to $p$.  As we
are interested in condensation-like phenomena, the simulations 
output graphs of the number of particles on the slow site with time.
Some typical examples of these graphs are shown in
Figure~\ref{3animouts}.

Of course, on a finite system, strictly one cannot identify
thermodynamic phases. However, it is useful to think of putative
phases emerging in large systems. In the following and
Figure~\ref{3animouts} `phase' is often taken to mean such a putative
phase.  

Inspecting Figure~\ref{3animouts} we see that the overshoot is
initially an extension of the fluid phase (Figure~\ref{3animouts}a);
up to the maximum of $v$ the system is still in the fluid phase. At
the maximum a condensate begins to emerge on the slow site.  In the
range of densities just beyond the maximum of the overshoot the
emergent condensate appears intermittently, but is not stable over
long time periods (Figure~\ref{3animouts}b).  We interpret this as
`coexistence' between the fluid and condensed phases, i.~e.\ at any
given time the system is in one or the other.  Eventually this
coexistence gives way to a clear condensate as $v$ drops back down to
its predicted maximum value of $p$ (Figure~\ref{3animouts}c).

Having established what takes place in the overshoot region,
we proceed in the next section to give an analytical account for this effect
working within the canonical ensemble.

\section{Canonical analysis}
\label{Sec:Canon}
The SDS detailed in (\ref{simplehop1},
\ref{simplehopmu}), has a convenient property in
that the canonical normalisation, $\mathrm{Z}(M,L)$
(\ref{bigZcanongeneral}), can be reduced to a single sum.  This makes
it possible to analyse the model directly within the canonical
ensemble.

The form that $\mathrm{Z}(M,L)$ takes for the SDS model is
\begin{equation}
\mathrm{Z}(M,L) = \sum_{n=0}^{L} p^{-n} \frac{n!}{(1+\lambda)_{n}} 
\binom{L+M-n-2}{M-2}\;.
\label{Zcanon}
\end{equation}
Here we have simply inserted the $\mathrm{f}_{\mu}$
(\ref{simplef1},\ref{simplefmu}) into the general expression for
$\mathrm{Z}(M,L)$ (\ref{bigZcanongeneral}).  The simple form of the
hopping rates for sites $\mu > 1$ 
(\ref{simplehopmu})
allows the sum over $n_{2} \dots
n_{M}$ to be performed easily, yielding the combinatoric factor which
comes from the delta function
in (\ref{bigZcanongeneral}).  We also re-label $n_{1}$ as $n$ for
convenience.

The normalisation is the sum over all un-normalised probabilities of
occupancies of the sites.  This means that the magnitude of the
$n^{\mathrm{th}}$ term in the sum (\ref{Zcanon}) is proportional to
the probability of the slow site having occupancy $n$.  Thus if
$\mathrm{Z}(M,L)$ is dominated by a maximum at low $n$, the
corresponding system is in the fluid phase.  Conversely, if
$\mathrm{Z}(M,L)$ is dominated by a maximum at large $n$, the
corresponding system is in the condensed phase.  If there are two
maxima of similar magnitudes 
we interpret this as phase coexistence.  
Thus investigating the maxima of this sum should furnish us
with an explanation of the overshoot.

To investigate the maxima of this sum, we look for its stationary
points, i.~e.\ points where the ratio of two consecutive terms tends
to one.  Solving for these points gives a quadratic equation in $n$,
the solutions of which are:
\begin{multline}
n =  \frac{M}{2} \left[ \phi - \phi_{\mathrm{c}}  \right] + 
\frac{A}{2} 
{}\pm  \frac{M}{2} \left[ \left( \phi - \phi_{\mathrm{c}}  \right)^{2} + 
\frac{1}{M} \left\{ 2 \left( \phi - \phi_{\mathrm{c}}  \right) A - 
4 \lambda \phi_{\mathrm{c}} (1+\phi) \right\} \right. \\ 
\left. {} + \frac{1}{M^{2}} \left\{ 
A^{2} 
+ 4 \lambda \phi_{\mathrm{c}} \right\} \right]^{\frac{1}{2}}\;,
\label{nturningpts}
\end{multline}
where $\phi_{\mathrm{c}} = p/(1-p)$ is the expected critical density
and the constant $A = \phi_{\mathrm{c}}/p +
\phi_{\mathrm{c}}(1+\lambda)$.

The nature of these turning points can be determined by considering the
ratio of the first two terms and also of the last two
in (\ref{Zcanon}).  In the
following we also assume that $p \leq 1/(1+\lambda)$ for simplicity,
although this turns out to make little difference to the overall
picture.

\begin{figure}[htb]
\begin{center}
\includegraphics[angle=0,width=10cm]{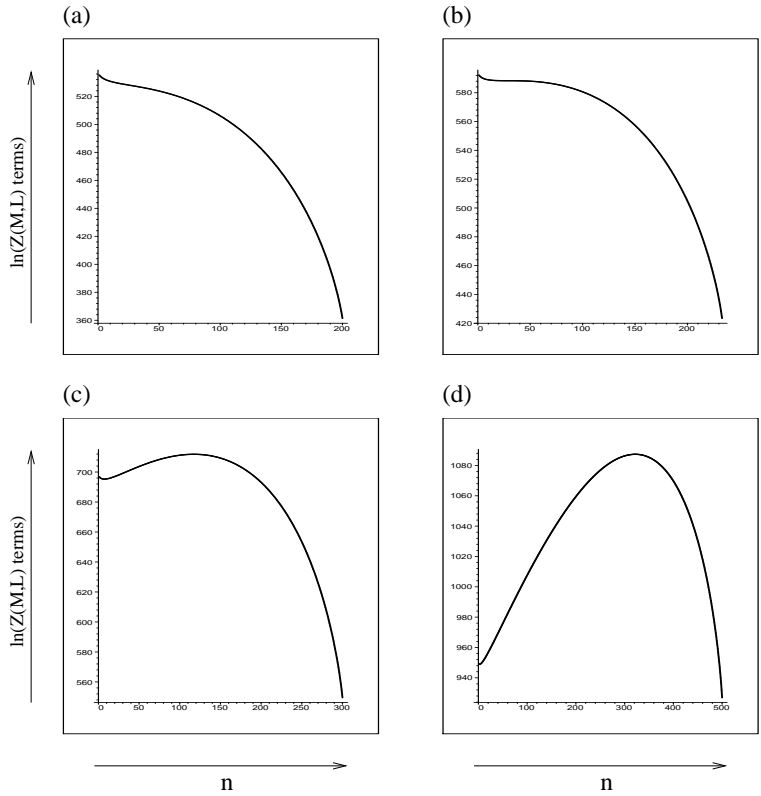}
\caption{
Graphs showing the logarithm of the terms from the normalisation sum
$\mathrm{Z}(M,L)$ (\ref{Zcanon}), for a system of size $M=1000$, with
$\lambda = 4$ and $p=0.15$. (a) $L=200$ ($\phi<\phi_2$), boundary
maximum at $n=0$, no turning points. (b) $L=233$ ($\phi = \phi_2$),
boundary maximum at $n=0$ and stationary point at higher $n$. (c)
$L=300$ ($\phi > \phi_2$), boundary maximum at $n=0$ with greater
maximum at high $n$. (d) $L=500$ ($\phi \gg \phi_2$), boundary minimum
at $n=0$ and maximum at high $n$. \label{turningptsfig}}
\end{center}
\end{figure}

First we note that the roots in
(\ref{nturningpts}) become real at a density $\phi_2$ given by
\begin{equation}
\phi_{2} = \frac{p}{1-p} - \frac{1}{M} \left( \frac{1+p(1-\lambda)}{1-p} \right) +
\left[ \frac{1}{M} \frac{4 \lambda p}{(1-p)^{2}} - 
\frac{1}{M^{2}} \frac{8 \lambda p}{(1-p)^{2}} \right]^{\frac{1}{2}} \;.
\label{ph2}
\end{equation}
Thus on increasing the density from zero the profile of the terms in
the sum goes through the sequence illustrated in
Figure~\ref{turningptsfig}: (a) For $\phi < \phi_2$ there is a
boundary maximum at $n=0$; (b) at $\phi =\phi_2$ a stationary point
emerges; (c) for $\phi >\phi_2$ a maximum emerges at large $n$ in
addition to the local maximum at $n=0$; (d) the maximum at large $n$
increases and dominates, with $n=0$ eventually becoming a boundary
minimum.

We interpret this sequence as follows.  At low density, $\phi <
\phi_{2}$ there are no turning points in the physical region
$0 \leq n \leq L$, with $n$  real.  Thus the
values of $n$ near  $n=0$ dominate and this corresponds to the fluid phase.
If we were to consider the $M\to \infty$ limit of (\ref{ph2}) we would
obtain $\phi_2=\phi_{\mathrm{c}}$.  However for $M$ finite,  the
square root part of (\ref{nturningpts}) remains imaginary up to
$\phi_2 > \phi_{\mathrm{c}}$ and the system remains in the fluid phase
until this point.  Thus the maximum  at
high $n$ emerges  beyond the expected critical density $\phi_{\mathrm{c}} =
p / (1-p)$ by an amount ${\cal O}(M^{-1/2})$:
\begin{equation}
\phi_2 =
\frac{p}{1-p} + \frac{2}{1-p} \left( \frac{\lambda p}{M}\right)^{1/2}
+{\cal O}(\frac{1}{M})\;.
\end{equation}
When $\phi$ is just greater than  $\phi_{2}$ we interpret the
boundary maximum at $n=0$ and the turning point at large $n$ as
coexisting fluid and condensed phases. 
Finally, for $\phi$ much
larger than $\phi_2$ the maximum at high $n$ dominates and we
interpret this as the condensed phase.  

We expect that the maximum of the overshoot will be near to the point
$\phi_{2}$ as this is where a putative condensate can begin to form
and reduce the current.  We see that an amount ${\cal O}(M^{-1/2})$
beyond $\phi_{\mathrm{c}}$ is consistent with the data from exact
numerical calculation of the current (Figure~\ref{vdiffMfig}).  
In fact, we compared the peak
of the overshoot with $\phi_{2}$ for systems of size $M=1000$ we found
very good agreement for $\lambda \approx 1 \dots 6$.  Outwith this
range we believe that further finite size effects are becoming
important.

The form of $\mathrm{Z}(M,L)$, given in equation (\ref{Zcanon}), also
allows the behaviour of the current in the condensed phase to be
analysed.  One finds that the asymptotic behaviour of the current for
large particle number (i.~e.\ where the sum in (\ref{Zcanon}) is
dominated by a maximum at large $n$) is
\begin{equation}
v = p \left[ 1 + \frac{1}{M} \frac{\lambda}{(\phi-
\phi_\mathrm{c})} + {\cal O} \left( \frac{1}{M^{2}} \right) \right]\;.
\label{vovercanon}
\end{equation}
Thus for $\lambda >0$ the current approaches its asymptotic value from
above as $\phi \to \infty$.  It should be noted that the second term
in the expansion is small only for $\phi - \phi_{\mathrm{c}} \gg
\lambda/M$.

\section{Discussion}
In this work we have studied the condensation transitions of a ZRP
with a single defect site, in both the grand canonical and canonical
ensembles.  Analysis in the grand canonical ensemble predicts the two
different condensation mechanisms (A and B) discussed in
Section~\ref{mechsec}. For $\lambda >2$ mechanism A applies which
predicts a finite size overshoot in the current.
The approach of $v$ to
its saturation value $p$ is $v/p = 1+{\cal O}(\ln L/L)$.
For  $\lambda \leq 2$
mechanism B applies and there should be no overshoot.  

However, simulations in the canonical ensemble reveal an overshoot for
all $\lambda >0$; only for $\lambda =0$ do we see the expected
mechanism B behaviour. Analysis within the canonical ensemble confirms
this and predicts that for $\lambda >0$ the overshoot in the current
peaks at a density $\phi=\phi_{\mathrm{c}} + {\cal O}(M^{-1/2})$.  The
approach of $v$ to its saturation value as particle number increases
is $v/p = 1+{\cal O}(1/L)$.  Thus it is not clear to what extent the
mechanisms A and B are relevant or well-defined within the canonical
ensemble.  Moreover, from the point of view of finite size critical
behaviour, the two ensembles are not equivalent.

It would be interesting to further understand the finite size
inequivalence of the ensembles.  In order to shed some light on this
one can follow a standard approach \cite{khuangbook,BBJ,BrazJPhysMRE}
to evaluating the canonical normalisation (\ref{bigZcanongeneral}) in
terms of the grand canonical partition function (\ref{Zgce}).  We use
the integral representation of the delta function to write the
canonical normalisation as a contour integral in the complex plane
\begin{equation}
\mathrm{Z}(M,L) = \oint \frac{\mathrm{d}s}{2 \pi i} s^{-L+1}
{\cal Z}(s)\;.
\label{Zint}
\end{equation}
Then for large system size and particle number, $M$ and $L$ respectively,
this integral will be dominated by its saddle point, given by
\begin{equation}
\phi = \frac{s}{M} \sum_{\mu = 1}^{M} 
\frac{\mathrm{F}_{\mu}^{\prime}(s)}{\mathrm{F}_{\mu}(s)}\;,
\label{sad}
\end{equation}
where $\phi$ is the particle density ($L/M$).  This equation precisely
recovers (\ref{saddlepoint1}). However it must be borne in mind that
(\ref{sad}) holds only when the saddle point of (\ref{Zint}) is {\em
valid}, i.~e.\ when it dominates the integral (\ref{Zint}).  It would
not be surprising for the saddle point to break down when it
approaches the radius of convergence of ${\cal Z}(s)$ in a way that
depends on system size, as in the analysis of (\ref{saddlepoint1}) in
section~\ref{CondGCEsec}.  Further analysis of this
phenomenon would be enlightening.

Finally it would be of interest to perform simulations within the
grand canonical ensemble to confirm our analytical predictions. To do
this one would introduce transitions whereby particles are created and
annihilated with appropriate rates.

\begin{acknowledgments}
We thank M. E. Cates for useful discussions and comments.
A. G. A. thanks the Carnegie Trust for the Universities of Scotland
for a studentship.  D. M. is supported by the Israel Science
Foundation (ISF) and the Einstein Center and visits to Edinburgh were
funded by EPSRC under grant number GR/R52497.
\end{acknowledgments}

\end{document}